# Sub-6GHz Assisted MAC for Millimeter Wave Vehicular Communications

Baldomero Coll-Perales, Javier Gozalvez, and Marco Gruteser

*Abstract*— Future connected and automated driving applications can require larger bandwidth and higher data rates than currently supported by sub-6GHz V2X technologies (e.g. DSRC, ITS-G5 or C-V2X). This has triggered the interest in developing mmWave vehicular communications. However, solutions are necessary to solve the challenges resulting from the use of high-frequency bands and the high mobility of vehicles. This paper contributes to this active research area by proposing a sub-6GHz assisted mmWave MAC that decouples the mmWave data and control planes. The proposal offloads mmWave MAC control functions to a sub-6GHz V2X technology. This approach improves the operation of the MAC as the control functions benefit from the longer range, and the broadcast and omnidirectional transmissions of sub-6GHz V2X technologies. This study demonstrates that the proposed sub-6GHz assisted mmWave MAC reduces the control overhead and delay, and increases the spatial sharing compared to mmWave communications using a configuration of IEEE 802.11ad tailored to vehicular networks. The proposed MAC is here evaluated for V2V communications using 802.11p for the control plane and 802.11ad for the data plane, although it can be adapted to other technologies such as C-V2X and 5G NR-V2X.

## I. INTRODUCTION

VEHICULAR networks will support the exchange of information between vehicles (Vehicle to Vehicle, V2V), and between vehicles and other nodes (V2X, Vehicle to Everything). The V2X standards ITS-G5, DSRC and ITS Connect are based on the 802.11p amendment to the IEEE 802.11 standard (amendment 6: Wireless Access in Vehicular Environments), and operate on the 5.9GHz or 760MHz bands. The 3GPP has also developed an adaptation of LTE to support sub-6GHz V2X communications known as C-V2X or LTE-V [1]. Sub-6GHz V2X technologies have been designed to support active safety services that require low data rates broadcast communications.

The communication and bandwidth requirements of connected and automated driving applications (Section II.A) can challenge existing sub-6GHz V2X standards with limited bandwidth and data rates. This has motivated studies to design novel V2X technologies that will not replace but rather complement existing sub-6GHz ones [2]. In particular, studies have been initiated to investigate the potential of utilizing millimeter wave (mmWave) communications for connected and automated driving. MmWave provides significantly larger bandwidth than sub-6 GHz technologies, but is more vulnerable to blockage and suffers from higher pathloss. Highly directional antennas can increase the link budget in a beam, but the highly dynamic vehicular environment presents many challenges to the beamforming and tracking processes, as well as to the design of MAC (Medium Access Control) protocols for scheduling mmWave vehicular communications.

The requirements of connected and automated driving use cases, and the different characteristics of sub-6GHz and mmWave V2X communications, will require leveraging multiple radio access technologies for V2X communications [3]. Sub-6GHz V2X technologies could support basic broadcast safety services, while mmWave and NR (New Radio) V2X could be used for enhanced use cases (eV2X) demanding higher data rates. The envisioned multi-band, multi-link and multi-technology V2X scenario offers opportunities to address some of the challenges experienced by mmWave in high mobility vehicular scenarios. For example, [4] proposes the use of side (or out-of-band) information to obtain the relative position of vehicles, and reduce the beam alignment overhead of mmWave V2I communications. This information can come from automotive sensors (e.g. LIDAR, cameras) or DSRC. [5] further reduces this overhead by exploiting similarities between sub-6GHz and mmWave channel parameters (e.g. spatial characteristics of the signals).

This study builds on the foreseen multi-band, multi-link and multi-technology V2X scenario, and goes a step beyond the current state-of-the-art by proposing to exploit sub-6GHz V2X communications to design a MAC for mmWave vehicular communications. In particular, this study proposes to decouple the control and data planes, and use sub-6GHz broadcast omnidirectional V2X communications for the control plane and directional mmWave communications for the data plane. Our proposal does not modify the MAC (or PHY) of sub-6GHz V2X technologies, but exploits their larger communications range to facilitate the mmWave beam alignment, identify links and neighbors, and schedule mmWave data transmissions. The

B. Coll-Perales and J. Gozalvez are with Universidad Miguel Hernández de Elche, Spain. M. Gruteser is with Rutgers University, N.J, USA.

obtained simulation results demonstrate that the proposed MAC significantly improves the communications performance, and reduces the overhead compared to a IEEE 802.11ad mmWave MAC tailored for vehicular communications.

## II. MMWAVE VEHICULAR COMMUNICATIONS: STATUS AND CHALLENGES

### A. Connected and automated driving

Connected and automated driving eV2X applications will require vehicles to exchange additional messages to the beacons transmitted using sub-6GHz V2X technologies (referred to as BSMs -Basic Safety Messages- or CAMs -Cooperative Awareness Message). These messages can be significantly larger in size than the beacons, and include information such as sensor data, detected objects, or the vehicles' planned and desired trajectory. The technical community is currently studying with what frequency this information should be transmitted and what should its resolution/accuracy be. For example, 3GPP specifies in [3] the 'collective perception of environment' use case that will require vehicles to exchange ~1600-byte messages and support data rates between 50Mbps and 1Gbps for the transmission of low-resolution/pre-processed and high-resolution/raw sensor data, respectively. According to the 3GPP, the use case 'information sharing for partial/conditional automated driving' will require the transmission of messages of up to 6500bytes with a rate of 10Hz (i.e. 0.52Mbps). This size is estimated considering 100 detected objects and 65bytes per object. This size is in line with the message format defined by ETSI in [6]. The use case 'information sharing for high/full automated driving' requires vehicles to share high resolution perception data at 50Mbps. The 50Mbps requirement takes into account: ~10Mbps (HD camera) + ~35Mbps (LIDAR with 6 vertical angles, 64 vertical elements, 10Hz horizontal rotation) + other sensor data. These data rate requirements are per vehicle, so the total bandwidth needed in an area increases with the number of vehicles. These enhanced V2X use cases require data rates and bandwidth levels that cannot be supported by sub-6GHz V2X technologies; the maximum data rate in practice of 802.11p is below 10Mbps [4].

### B. Standardization

The described use cases highlight the need to develop new V2X standards that can support larger data rates. To this aim, 3GPP started under Release 15 a study item on 'Enhancement of 3GPP Support for V2X scenarios' [3] [7]. The study item has identified the need of a 5G New Radio (NR) operating above-6GHz (including the mmWave band) to support the use cases described in Section II.A; a first non-standalone 5G NR using an LTE anchor for the control plane was actually approved in December 2017. The study item also advocates for the support of V2X services using multiple radio access technologies including NR and sub-6GHz V2X technologies. 3GPP indicates that 5G NR will complement and not replace sub-6GHz V2X technologies [2].

Another option for mmWave V2X communications is the IEEE 802.11ad standard that can support data rates up to 7Gbps. Similarly to 802.11a and 802.11p, 802.11ad could serve as a basis for developing a 802.11-based mmWave V2X standard [4][8]. A first evaluation of the 802.11ad MAC for V2V communications was presented in [8]. This study highlights important MAC inefficiencies of the current IEEE 802.11ad standard when utilized for V2V communications. In particular, [8] shows that the 802.11ad MAC processes for neighbor identification, beamforming and scheduling, generate significant overhead under vehicular scenarios. In addition, the current 802.11ad MAC results in many scheduling conflicts and coordination problems between different 802.11ad vehicular transmitters that significantly degrade the network performance. These findings have motivated this study, and the need to design a novel mmWave MAC that can efficiently support mmWave V2X communications.

### C. PHY features

MmWave communications can be subject to severe propagation pathloss, and require forming narrow beams with high antenna gains between transmitters and receivers in order to increase the link budget. The use of multiple antennas and directional beams leads to several physical layer (PHY) challenges. One of them is the beamforming or transceiver architecture design that can be analog, digital or hybrid [9]. Another one is the selection and impact of the beamwidth [10]. Using narrow beams augments the antenna gain, reduces the signal's multipath components and the Doppler spread, and limits the interference area. However, it can increase the complexity and overhead to align the transmitter and receiver beams. Vehicles need to align their beams to increase the link budget. Misaligned beams can result in a link outage (a.k.a. deafness problem). Standards such as 802.11ad define the processes to align beams. They rely on a handshaking process between transmitter and receivers across each beam; the process is repeated sequentially for each beam. These processes can be particularly challenging in highly mobile vehicular scenarios, and generate significant overhead [8].

### D. MAC challenges

Fig. 1 illustrates some of the MAC challenges of mmWave vehicular communications due to its distinctive PHY features. Without loss of generality, Fig. 1 focuses on V2V and considers an analog beamforming architecture.

- *Link availability and identification of neighbors.* MmWave links are easily blocked by obstacles, e.g. buildings or vehicles (Fig. 1.a). The abundance of obstructions makes uncertain the availability of mmWave links, and a challenge is hence identifying the neighbors under LOS (Line of Sight) conditions and high link budgets.
- *Quasi-omni and directional carrier sensing.* The use of

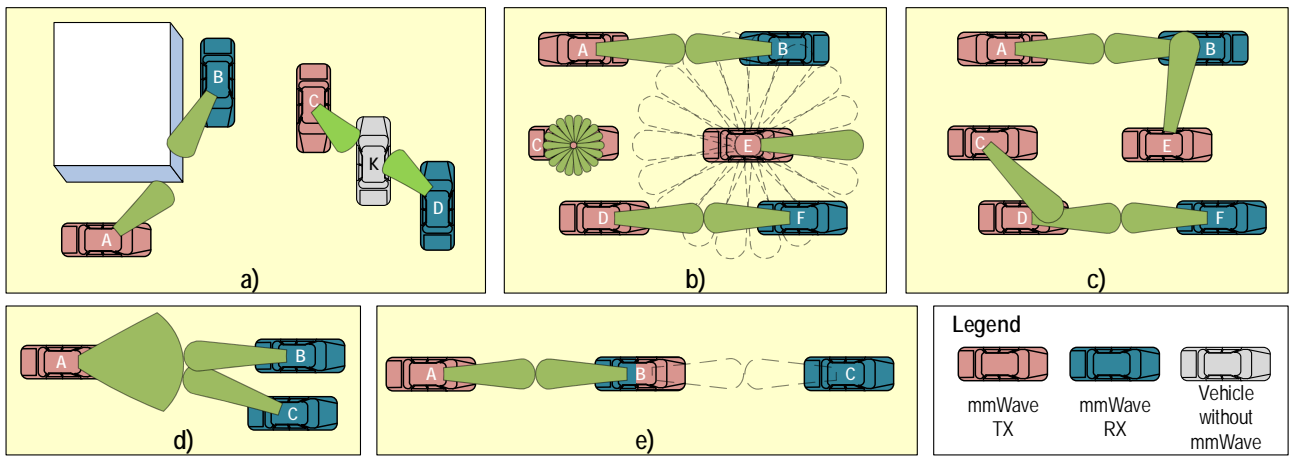

Fig. 1. MAC challenges of mmWave vehicular communications: a) Link availability and identification of neighbors; b) Quasi-omni and directional carrier sensing; c) Scheduling; d) Beamwidth-aware scheduling; e) Relaying.

directional beams in mmWave limits to a sector the area that can be sensed by vehicles. The array of mmWave antennas can be configured to form a quasi-omnidirectional pattern at the expense of a lower antenna gain. In this case, vehicles can sense other vehicles in all directions, but the range is significantly reduced compared to directional beams. Fig. 1.b illustrates an example in which the A-B and D-F mmWave links are active. Vehicles C and E want to communicate with their neighbors, but their antenna configuration (quasi-omni and directional, respectively) does not allow them to identify/sense whether their neighbors have active mmWave links or not.

- *Scheduling*. V2V usually requires the use of distributed scheduling schemes. These schemes generally rely on the sensing capabilities of the vehicles to coordinate the access to the medium. For example, 802.11p uses CSMA/CA, and the vehicles access the channel when they sense it is idle for a while. The distributed scheduling in C-V2X mode 4 [1] also uses a sensing scheme for vehicles to identify which radio resources are not being used by other vehicles. The use of sensing-based scheduling in mmWave vehicular communications is challenging. Fig. 1.c represents the same scenario as Fig. 1.b. In this case, vehicle C cannot detect the transmissions from vehicles A and D due to its reduced sensing range (Fig. 1.b). It then considers the wireless medium as idle and starts a mmWave transmission to D. However, its data will not be received by D that is busy transmitting to F. Vehicle E does not detect the active A-B link due to its directional sensing range (Fig. 1.b), and starts a mmWave transmission to B. The transmissions from A and E collide at B. These examples illustrate the need for alternative distributed mmWave scheduling schemes. These schemes should also take advantage of directional beams to support multiple simultaneous transmissions between different pairs (referred to as spatial sharing or reuse).

- *Beamwidth-aware scheduling*. MmWave may trade-off beamwidth (and antenna gain) for coverage area at no reliability cost if vehicles are close and under LOS. This flexibility could be exploited to schedule a mmWave transmission to several receivers at the same time by configuring the beamwidth (Fig. 1.d). In this case, the challenges include: identifying the proximity of the receivers; adjusting the beamwidth so that the intended receivers can be addressed simultaneously while the antenna gain is sufficient to guarantee a reliable transmission; and integrating the adjustment of the beamwidth into the scheduling mechanism.

- *Relaying*. Relaying could help overcome the mmWave link budget and blockage challenges, and extend the coverage range. In this case, a tight coordination between the mmWave scheduling and relaying processes is needed as illustrated in Fig. 1.e. In the example, such coordination is needed to decide when vehicle B should be configured as receiver for the A-B link (1st hop), and when it should be configured as transmitter for the B-C link (2nd hop).

## III. SUB-6GHZ ASSISTED MAC FOR MMWAVE VEHICULAR COMMUNICATIONS

This work proposes a sub-6GHz assisted mmWave MAC designed to address some of the challenges discussed in Section II. The proposed MAC decouples the mmWave data and control planes, and offloads mmWave control functions to sub-6GHz V2X technologies such as DSRC, ITS-G5 or C-V2X. This work proposes to exploit the longer range, and broadcast and omnidirectional transmissions of sub-6GHz V2X to improve the operation of the mmWave MAC. In particular, the proposed scheme offloads the beamforming, link availability identification, and scheduling mmWave control functions to the sub-6GHz band. Without loss of generality, this study focuses on mmWave V2V communications.

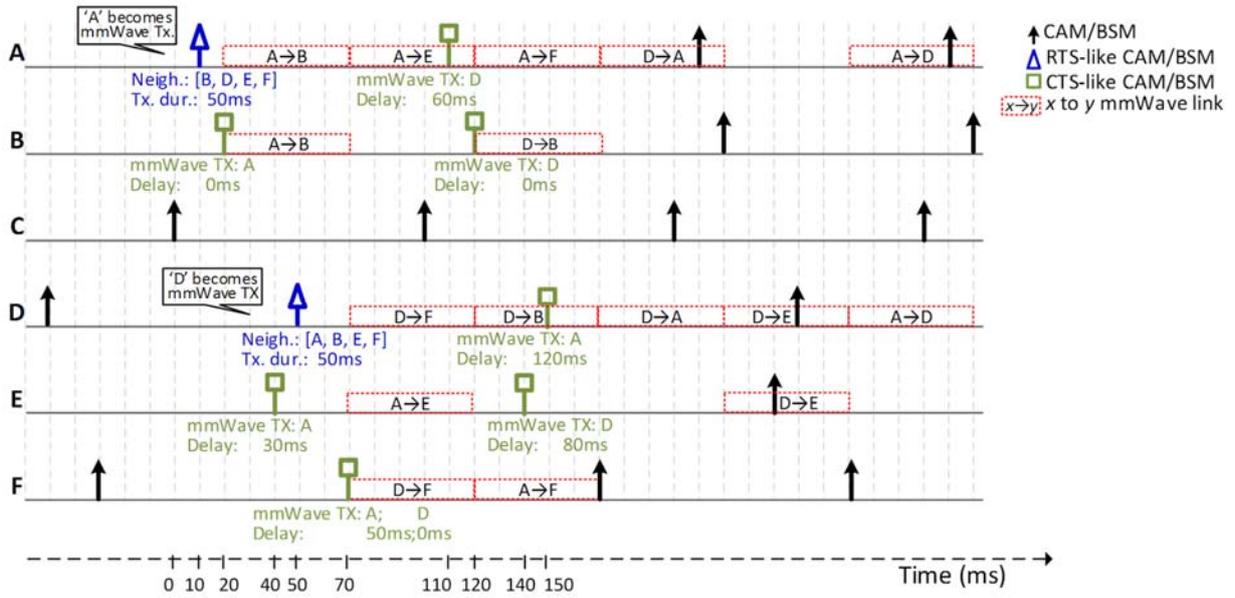

Fig. 2. Sub-6GHz assisted mmWave scheduling.

*A. Sub-6GHz assisted mmWave beamforming and link availability identification*

Following the proposal in [4] for V2I, this work uses sub-6GHz V2V communications to support the mmWave V2V beam alignment process. In particular, our implementation utilizes the status information (location, speed, acceleration and heading direction) transmitted in the sub-6GHz beacons to identify the location of neighboring vehicles. MmWave transmitter and receivers use the location information to select the beams that point towards each other. This significantly reduces the overhead of the beam alignment defined in 802.11ad. We also utilize sub-6GHz beacons to identify available links, i.e. neighboring vehicles under LOS conditions. To this aim, vehicles use the information transmitted in the beacons, in particular, the vehicles' location and dimensions.

*B. Sub-6GHz assisted mmWave scheduling*

This work proposes to leverage the transmission of sub-6GHz V2X beacons to schedule mmWave transmissions. The proposed mmWave scheduling exploits sub-6GHz V2X features to: 1) provide a contention-free (scheduled) access to the mmWave channel that does not require sensing; 2) minimize the control overhead; 3) maximize the mmWave channel utilization for data transmissions; and 4) improve the spatial sharing. To this aim, the scheduling scheme uses the transmission of sub-6GHz V2X beacons to announce scheduling decisions and organize the access to the mmWave channel. This is done without modifying the regular generation of beacons, hence guaranteeing that the proposed scheduling does not affect the normal operation of vehicular applications relying on sub-6GHz V2X communications. In addition, our scheduling proposal benefits from the reliability of sub-6GHz transmissions under LOS [11].

Fig. 2 illustrates the proposed scheduling scheme. CAMs/BSMs are periodically transmitted in Fig. 2, although the proposed scheme works with non-periodic transmissions. The vertical arrows represent the sub-6GHz V2V beacons. At t=0ms, vehicle A wants to transmit data using its mmWave interface. After A detects its available links, it uses its next beacon (t=10ms) to announce the mmWave neighbors it wants to communicate with (B, D, E and F), and the duration of its transmission to each neighbor (Tx. dur.=50ms). This information is attached to the beacon. The extended beacon acts as a Request-To-Send (RTS-like CAM/BSM in Fig. 2) for the addressed mmWave neighbors.

The scheduling of mmWave transmissions is decided in a distributed manner by the addressed neighbors as follows. B is the first addressed neighbor that transmits a beacon after the RTS-like CAM/BSM message from A. B uses this beacon as a Clear-To-Send to A (CTS-like CAM/BSM in Fig. 2). B attaches to the beacon the ID of the mmWave transmitter (A), and the time at which the transmission from A to B should start. Since B is the first vehicle responding to the request from A, this time is set equal to 0ms by B (delay=0ms). RTS-like and CTS-like CAM/BSMs are regular beacons, so all vehicles in the communications range are aware of the scheduling indications. The transmission from A to B starts as soon as A receives the CTS-like CAM/BSM from B. Fig. 2 shows the 50ms interval allocated for the A→B transmission at A and B. At t=40ms, E uses its next beacon as a CTS for A. E overheard the CTS-like CAM/BSM from B. Then, E indicates in its CTS-like CAM/BSM that the A→E transmission should be delayed 30ms. The A→E mmWave transmission starts as soon as the A→B transmission ends. D and F follow a similar process to schedule their transmissions without any conflict. However, in the scenario depicted in Fig. 2, D also becomes a mmWave transmitter and transmits at t=50ms a RTS-like CAM/BSM to announce the neighbors it wants to communicate with. D then postpones its CTS-like CAM/BSM replying to A's request until t=150ms (in its next beacon). F is in the list of mmWave neighbors of A and D. F uses its beacon at t=70ms for replying to both vehicles considering the messages F has previously overheard. In particular, F detects that its communication with A cannot be scheduled until the A→E transmission ends (delay=50ms), but the D→F transmission can be immediately scheduled (delay=0ms). The example illustrates how the proposed scheduling supports spatial sharing: for example, A→E and D→F transmissions happen simultaneously from t=70ms to t=120ms.

## IV. EVALUATION

The performance of the proposed scheme is compared against a reference optimum configuration of the 802.11ad MAC for V2V communications derived in [8]. The channel is divided in 802.11ad into cyclic intervals. Each interval is divided into a control and data interval. A mmWave transmitter uses the control interval (Beacon Header Interval in 802.11ad) to identify its neighbors, align beams, and schedule its transmissions. It uses the data interval (Data Transmission Interval in 802.11ad) to exchange data frames with the scheduled neighbors. Vehicles implementing our proposed MAC use 802.11p for their control plane and 802.11ad for their data plane. The mmWave channel is hence only used for exchanging data frames, and all mmWave control functions are offloaded to 802.11p. Vehicles transmit beacons on 802.11p every 100ms using an omnidirectional antenna, a transmission power of 15dBm, and the 6Mbps data rate. The data plane uses 802.11ad with an analog beamforming architecture and a 14-sector antenna, a transmission power of 10dBm, and a data rate of 693Mbps. This configuration is also used in the reference 802.11ad MAC that is configured following [8].

This study is conducted using ns-3.26, and leveraging the 802.11ad implementation in [12]. Additional features necessary to simulate the proposed MAC have been added. The evaluation is conducted under a highway scenario with 4 lanes and a vehicular density of 125vehicles/Km. In this scenario, each vehicle has on average 5.5 neighboring vehicles under LOS. The same scenario was implemented in [8] to evaluate 802.11ad for V2V communications. Following [8], the 802.11ad MAC is configured to accommodate 5 neighbors, and the duration of the interval for the data exchange with each neighbor is set to 50ms. For a fair comparison, the proposed sub-6GHz assisted mmWave MAC is evaluated considering that a mmWave transmission between two vehicles also lasts 50ms. During the 50ms data exchange interval, a mmWave transmitter sends 600 packets of 1600bytes each to a neighboring vehicle. The packet size is set following the 'collective perception of environment' use case (Section II.A) [3]. Simulations are conducted for different ratios of mmWave transmitters in the scenario ($R_{TX}$): {15, 20, 25, 30, 35, 40}%. In this study, we consider that each mmWave transmitter wants to communicate with all its neighboring vehicles under LOS conditions (including other mmWave transmitters). Multiple simulation runs have been executed to ensure the statistical accuracy of the results. The worst-case margin of error for the average results is below 4.5% with 95% confidence intervals. Simulations have also been conducted for a vehicular density of 250vehicles/Km that show similar trends to those reported here.

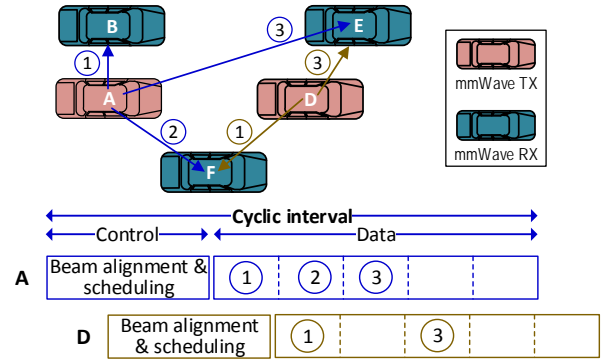

Fig. 3. Scheduling conflicts in IEEE 802.11ad V2V communications.

### A. Scheduling conflicts

This section analyzes the capacity of the schemes under evaluation to schedule the transmissions from a mmWave transmitter to all its neighboring vehicles. This is estimated by means of the average ratio of neighboring vehicles under LOS that are scheduled for mmWave transmission to the total amount of neighboring vehicles under LOS. The reference 802.11ad MAC implementation results in that on average each 802.11ad transmitter is able to schedule its transmissions to {70, 67, 63, 58, 54, 41}% of its neighboring vehicles under LOS when $R_{TX}$ = {15, 20, 25, 30, 35, 40}%. All neighboring vehicles under LOS are not scheduled due to the lack of coordination between 802.11ad transmitters that results in multiple scheduling conflicts. Fig. 3 shows an example of such conflicts in 802.11ad considering the scenario used in Fig. 2. A is the first vehicle that becomes a transmitter. During the control interval, it aligns beams and schedules its transmissions. D's control interval coincides in time with A's control and data intervals. As a result, A cannot detect D (and vice versa), and only allocates slots for transmission (in the data interval) to vehicles B, F and E. This also results in that D cannot detect B, since B has an active link with A during D's control interval. D hence only schedules transmissions to F and E. Fig. 3 illustrates another scheduling conflict in 802.11ad: A and D allocate slots (2nd and 1st respectively) to F that coincide in time.

The proposed sub-6GHz assisted MAC solves the 802.11ad scheduling conflicts by offloading mmWave control functions to sub-6GHz V2X technologies. The conducted simulations showed that the proposed MAC avoids the 802.11ad scheduling conflicts, and vehicles can detect, schedule and communicate with 100% of their neighbors for all $R_{TX}$ values.

### B. Control overhead

The reference 802.11ad MAC results in that each 802.11ad transmitter needs to send approximately 5.800 control bytes to align its beam and schedule its transmissions to each one of its neighbors [8]. This includes all messages and handshaking needed to identify the neighbors, align beams and schedule the transmissions. The proposed mmWave MAC only requires for control the extra bytes added in the sub-6GHz V2X beacons for scheduling the mmWave transmissions. In the simulated scenario, this is equivalent to approximately 100 control bytes[1], which represents a 98% reduction of the control overheard compared to the reference 802.11ad MAC.

In this study, the duration of the 802.11ad control and data intervals is set to 35.84ms and 250ms, respectively [8]. 802.11ad then uses the mmWave channel for control functions during 12% of the time. The proposed sub-6GHz assisted mmWave MAC offloads all control functions to the sub-6GHz V2X channel. The mmWave channel is hence fully utilized for data transmissions. We have estimated the increase in the CBR (Channel Busy Ratio) resulting from the bytes added to the sub-6GHz beacons. The CBR represents the proportion of the time that the sub-6GHz channel is sensed as busy. The CBR is here estimated as explained in [11]. The conducted evaluation has shown that the proposed sub-6GHz assisted mmWave MAC only increases the CBR between 0.77% ($R_{TX}$ =15%) and 2.04% ($R_{TX}$ =40%).

### C. Delay

Fig. 4 represents the delay between the moment a vehicle becomes a mmWave transmitter and the moment at which it starts its data transmissions to the five scheduled neighbors in the scenario. For 802.11ad, the delay to the start of the first scheduled data transmission is equal to the time between the start of a cyclic interval and the start of the first allocated data slot (Fig. 3). The scheduling conflicts highlighted in Section IV.A result in that some slots are not allocated to any station. This explains why this delay is higher than the duration of the control interval (35.84ms). Fig. 4.b also shows that this delay increases with $R_{TX}$ due to the increasing number of scheduling conflicts.

The delay to the first scheduled mmWave data transmission is lower-bounded at 60ms for the proposed MAC. This is equal to the sum of the time elapsed since the vehicle becomes a mmWave transmitter to the transmission of its first RTS-like CAM/BSM message (on average, half a beacon period), and the time elapsed from the RTS-like message to the first CTS-like CAM/BSM message. The later time is on average 10ms for the simulated scenario. The delay to the first data transmission increases at a slower pace with $R_{TX}$ for the proposed MAC. For example, Fig. 4 shows that the proposed scheme reduces by 20% and 83% the delay to the first data transmission in comparison to 802.11ad when $R_{TX}$ is 15% and 40% respectively.

Neighbors are contacted sequentially after the previous transmission ends, and so the delay should increase by at least 50ms for each scheduled neighbor. The proposed MAC shows slightly larger values due the need to postpone mmWave transmissions in order to avoid scheduling conflicts (Fig. 2). However, significantly larger delay values are observed with the 802.11ad MAC. This is caused by multiple scheduling conflicts (Fig. 3) that result in that a transmitter requires several cyclic intervals to be able to communicate with all its neighbors.

### D. Spatial sharing

Fig. 5 compares the spatial sharing capabilities of the two evaluated MACs. The figure shows that the proposed MAC utilizes more efficiently the mmWave channel since it can schedule mmWave data transmissions for most of the time: 94% and 98% when $R_{TX}$ is 15% and 40% respectively. On the other hand, the 802.11ad MAC fails to schedule any data transmissions for 37% ($R_{TX}$ =15%) and 47% ($R_{TX}$ =40%) of the time. These differences are due to the multiple scheduling conflicts that are experienced with the reference 802.11ad MAC (Section IV.A). These conflicts increase with the number of mmWave transmitters since the 802.11ad MAC cannot adequately coordinate them. This is observed in Fig. 5 that shows that the 802.11ad MAC schedules multiple simultaneous data transmissions for a significantly smaller percentage of time compared to the proposed MAC. For example, the proposed MAC schedules two or more simultaneous mmWave data transmissions for 41% and 82% of the time when RTX is equal to 15% and 40%, respectively. These values decrease to 13% and 20% with the 802.11ad MAC. These results demonstrate that the proposed MAC is capable to schedule more mmWave data transmissions than the 802.11ad MAC, and hence utilize more efficiently the communications channel.

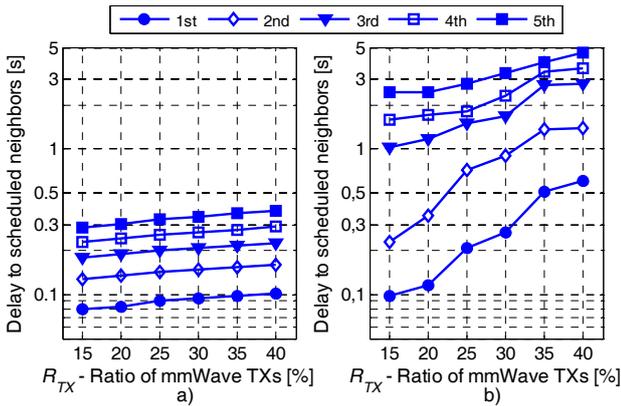

Fig. 4. Delay or time elapsed to the start of the mmWave data transmissions for the five scheduled neighbors in the scenario: a) Sub-6GHz assisted MAC, b) 802.11ad MAC.

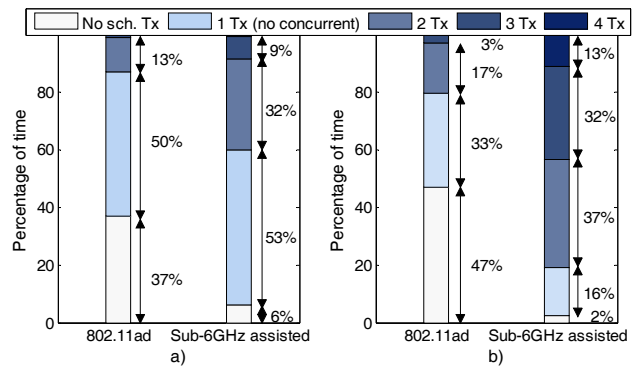

Fig. 5. Spatial sharing when $R_{TX}$ is 15% (a) and 40% (b). Percentage of time with no mmWave data transmissions scheduled (No sch. Tx) and percentage of time there are 1, 2, 3 or 4 mmWave data transmissions simultaneously scheduled.

---

[1] This includes the MAC addresses of neighbors (6 bytes), and the Tx. dur (2 bytes) and delay (2 bytes) fields of the RTS- and CTS-like CAM/BSM messages.

## V. Conclusions and Future Work

MmWave has been proposed to support connected and automated driving. However, the use of high-frequency bands and the mobility of vehicles create multiple challenges at the MAC level. This paper addresses these challenges with a novel sub-6GHz assisted MAC for mmWave vehicular communications. The proposal decouples the mmWave control and data planes, and offloads the mmWave MAC control functions (beamforming, link availability identification, and scheduling) to sub-6GHz V2X communications. The proposal exploits the longer range and the broadcast and omnidirectional transmissions of sub-6GHz V2X communications to improve the operation of the mmWave MAC. This study shows that the proposed MAC can solve important MAC challenges when using mmWave for V2V communications, and reduce the control overhead and delay compared to the IEEE 802.11ad standard. In addition, the proposed MAC increases the spatial sharing, and hence the network capacity and scalability.

The proposed sub-6GHz assisted mmWave MAC has been analyzed in this study using 802.11p in the control plane and 802.11ad in the data plane. However, the proposal is not restricted to these technologies, and can be adapted and extended to other technologies such as C-V2X and 5G NR. This is the case since C-V2X replaces the 802.11p PHY and MAC layers, but reutilizes the upper layers developed at ETSI, IEEE and SAE [13]. This includes the CAM/BSM messages that are utilized in our proposal to offload the mmWave control functions to the sub-6GHz V2X radio interface. Future extensions of our proposal could consider exploiting the beamwidth-aware scheduling and relaying capabilities to schedule several receivers at the same time, and reach neighbors at larger distances to the transmitter (even under NLOS). MmWave V2X communications can also exploit the directional beams to support multiple simultaneous transmissions between different pairs of vehicles. To this aim, it will be necessary to take into account the interference of active mmWave transmissions in the vicinity when scheduling the mmWave transmissions.

## Acknowledgment

B. Coll-Perales and J. Gozalvez acknowledge the support of the Spanish Ministry of Economy, Industry, and Competitiveness, AEI, and FEDER funds (TEC2017-88612-R, TEC2014-57146-R), and the Generalitat Valenciana (APOSTD/2016/049, AICO/2018/A/095).

## References

[1] R. Molina-Masegosa, J. Gozalvez, "LTE-V for Sidelink 5G V2X Vehicular Communications: A New 5G Technology for Short-Range Vehicle-to-Everything Communications", IEEE Veh. Technol. Mag., vol. 12, no. 4, 2017, pp. 30-39.
[2] 3GPP TR38.913 v15.0.0, "Study on scenarios and requirements for next generation access technologies", June 2018.
[3] 3GPP TR22.886 v16.1.1, "Study on enhancement of 3GPP Support for 5G V2X Services", Sept. 2018.
[4] J. Choi et al., "Millimeter Wave Vehicular Communications to Support Massive Automotive Sensing", IEEE Commun. Mag., vol. 54, no. 12, 2016, pp. 160-167.
[5] N. Gonzalez-Prelcic, et al., "Millimeter-Wave Communication with Out-of-Band Information", IEEE Commun. Mag., vol. 55, no. 12, 2017, pp. 140-146.
[6] ETSI TR103-562 v0.0.7, "Informative Report for the Collective Perception Service", Apr. 2018.
[7] 3GPP TR22.186 v15.2.0, "Service requirements for enhanced V2X scenarios", Sept. 2017.
[8] B. Coll-Perales, M. Gruteser, J. Gozalvez, "Evaluation of IEEE 802.11ad for mmWave V2V Communications", Proc. IEEE WCNCW, Barcelona, 2018, pp. 1-6.
[9] F. Sohrabi, W. Yu, "Hybrid digital and analog beamforming design for large-scale antenna arrays", IEEE J. Sel. Top. Signal Process., vol. 10, no. 3, 2016, pp. 501–513.
[10] C. Perfecto, J. Del Ser, M. Bennis, "On the interplay between scheduling interval and beamwidth selection for low-latency and reliable V2V mmWave communications", Proc. IEEE ICIN, Paris, 2017, pp. 1-8.
[11] M. Sepulcre, J. Gozalvez, B. Coll-Perales, "Why 6Mbps is not (always) the Optimum Data Rate for Beaconing in Vehicular Networks", IEEE T. Mobile Comput., vol. 16, no. 12, 2017, pp. 3568-3579.
[12] H. Assasa, J. Widmer, "Implementation and Evaluation of a WLAN IEEE 802.11ad Model in ns-3", Proc. ACM WNS3, Seattle, 2016, pp. 57-64.
[13] A. Papathanassiou, A. Khoryaev, "Cellular V2X as the essential enabler of superior global connected transportation services", IEEE 5G Tech. Focus, vol. 1, no. 2, 2017, June.
[14] W.-K. Chen, *Linear Networks and Systems*. Belmont, CA, USA: Wadsworth, 1993, pp. 123–135.

**Baldomero Coll Perales** (bcoll@umh.es) received the M.Sc.Eng. and Ph.D. degrees, both with honors, from the Universidad Miguel Hernandez (UMH) de Elche, Spain. He is currently Research Fellow at the UWICORE laboratory and Postdoctoral Associate at IIT-CNR (Pisa, Italy). He was formerly Postdoctoral Associate at WINLAB (Rutgers University, NJ, USA). His research interests lie in the field of advanced mobile and wireless communications systems, including device-centric technologies and V2X communications. He is Associate Editor for Springer's Telecommunication Systems and Int. J. of Sensor Networks. He has served as Track Co-Chair for IEEE VTC-Fall 2018, and as member of the TPC in over 25 international conferences.

**Javier Gozalvez** (j.gozalvez@umh.es) received an electronics engineering degree from the Engineering School ENSEIRB (Bordeaux, France), and a PhD in mobile communications from the University of Strathclyde, Glasgow, U.K. Since October 2002, he is with UMH, where he is Full Professor and Director of the UWICORE laboratory. At UWICORE, he leads research activities in the areas of vehicular networks, 5G and beyond, and industrial wireless networks. He is an elected member to the Board of Governors of the IEEE Vehicular Technology Society (IEEE VTS) since 2011, and served as its 2016-2017 President. He was an IEEE Distinguished Lecturer for the IEEE VTS, and currently serves as Distinguished Speaker. He is the Editor in Chief of the IEEE Vehicular Technology Magazine, and founded the IEEE Connected and Automated Vehicles Symposium.

**Marco Gruteser** (gruteser@winlab.rutgers.edu) received the M.S. and Ph.D. degrees from the University of Colorado, Boulder, USA, in 2000 and 2004, respectively. He is the Peter D. Cherasia Faculty Scholar and Professor of electrical and computer engineering as well as computer science (by courtesy) at the Wireless Information Network Laboratory at Rutgers University. He directs research in mobile computing, pioneered location privacy techniques, and has focused on connected vehicles challenges. He chairs ACM SIGMOBILE, has chaired conferences including ACM MobiSys and ACM MobiCom. He also held research and visiting positions at the IBM T. J. Watson Research Center, Carnegie Mellon University, and Google. His recognitions include an NSF CAREER award, a Rutgers Board of Trustees Research Fellowship for Scholarly Excellence and six award papers (including ACM MobiCom 2012, ACM MobiCom 2011 and ACM MobiSys 2010).